\documentclass[twoside,twocolumn,9pt]{article}
\usepackage{extsizes}
\usepackage[super,sort&compress,comma]{natbib} 
\usepackage[version=3]{mhchem}
\usepackage[left=1.5cm, right=1.5cm, top=1.785cm, bottom=2.0cm]{geometry}
\usepackage{balance}
\usepackage{times,mathptmx}
\usepackage{sectsty}
\usepackage{graphicx} 
\usepackage{lastpage}
\usepackage[format=plain,justification=justified,singlelinecheck=false,font={stretch=1.125,small,sf},labelfont=bf,labelsep=space]{caption}
\usepackage{float}
\usepackage{fancyhdr}
\usepackage{fnpos}
\usepackage[english]{babel}
\addto{\captionsenglish}{%
  
}
\usepackage{array}
\usepackage{droidsans}
\usepackage{charter}
\usepackage[T1]{fontenc}
\usepackage[usenames,dvipsnames]{xcolor}
\usepackage{setspace}
\usepackage[compact]{titlesec}
\usepackage{hyperref}
\usepackage[exponent-product=\cdot,per-mode=symbol]{siunitx}
\usepackage{epstopdf}%This line makes .eps figures into .pdf - please comment out if not required.

\definecolor{cream}{RGB}{222,217,201}

\begin{document}

\pagestyle{fancy}
\thispagestyle{plain}
\fancypagestyle{plain}{

%%HEADER%%%
\fancyhead[C]{\includegraphics[width=18.5cm]{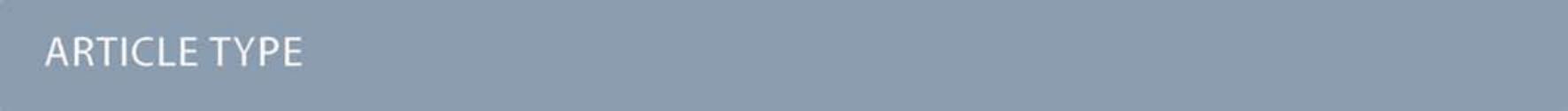}}
\fancyhead[L]{\hspace{0cm}\vspace{1.5cm}\includegraphics[height=30pt]{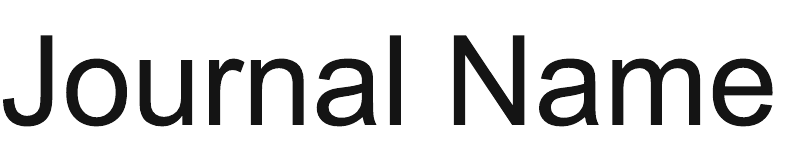}}
\fancyhead[R]{\hspace{0cm}\vspace{1.7cm}\includegraphics[height=55pt]{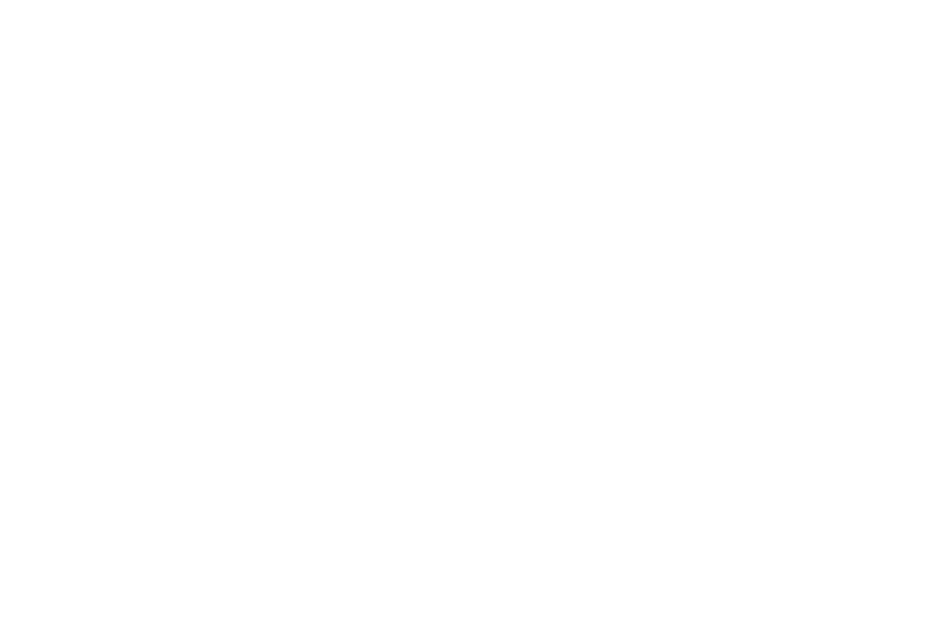}}
\renewcommand{\headrulewidth}{0pt}
}
%%%END OF HEADER%%%

%%%PAGE SETUP - Please do not change any commands within this section%%%
\makeFNbottom
\makeatletter
\renewcommand\LARGE{\@setfontsize\LARGE{15pt}{17}}
\renewcommand\Large{\@setfontsize\Large{12pt}{14}}
\renewcommand\large{\@setfontsize\large{10pt}{12}}
\renewcommand\footnotesize{\@setfontsize\footnotesize{7pt}{10}}
\makeatother

\renewcommand{\thefootnote}{\fnsymbol{footnote}}
\renewcommand\footnoterule{\vspace*{1pt}% 
\color{cream}\hrule width 3.5in height 0.4pt \color{black}\vspace*{5pt}} 
\setcounter{secnumdepth}{5}

\makeatletter 
\renewcommand\@biblabel[1]{#1}            
\renewcommand\@makefntext[1]% 
{\noindent\makebox[0pt][r]{\@thefnmark\,}#1}
\makeatother 
\renewcommand{\figurename}{\small{Fig.}~}
\sectionfont{\sffamily\Large}
\subsectionfont{\normalsize}
\subsubsectionfont{\bf}
\setstretch{1.125} %In particular, please do not alter this line.
\setlength{\skip\footins}{0.8cm}
\setlength{\footnotesep}{0.25cm}
\setlength{\jot}{10pt}
\titlespacing*{\section}{0pt}{4pt}{4pt}
\titlespacing*{\subsection}{0pt}{15pt}{1pt}
%%%END OF PAGE SETUP%%%

%%%FOOTER%%%
\fancyfoot{}
\fancyfoot[LO,RE]{\vspace{-7.1pt}\includegraphics[height=9pt]{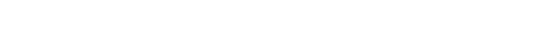}}
\fancyfoot[CO]{\vspace{-7.1pt}\hspace{13.2cm}\includegraphics{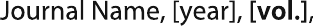}}
\fancyfoot[CE]{\vspace{-7.2pt}\hspace{-14.2cm}\includegraphics{RF}}
\fancyfoot[RO]{\footnotesize{\sffamily{1--\pageref{LastPage} ~\textbar  \hspace{2pt}\thepage}}}
\fancyfoot[LE]{\footnotesize{\sffamily{\thepage~\textbar\hspace{3.45cm} 1--\pageref{LastPage}}}}
\fancyhead{}
\renewcommand{\headrulewidth}{0pt} 
\renewcommand{\footrulewidth}{0pt}
\setlength{\arrayrulewidth}{1pt}
\setlength{\columnsep}{6.5mm}
\setlength\bibsep{1pt}
%%%END OF FOOTER%%%

%%%FIGURE SETUP - please do not change any commands within this section%%%
\makeatletter 
\newlength{\figrulesep} 
\setlength{\figrulesep}{0.5\textfloatsep} 

\newcommand{\topfigrule}{\vspace*{-1pt}% 
\noindent{\color{cream}\rule[-\figrulesep]{\columnwidth}{1.5pt}} }

\newcommand{\botfigrule}{\vspace*{-2pt}% 
\noindent{\color{cream}\rule[\figrulesep]{\columnwidth}{1.5pt}} }

\newcommand{\dblfigrule}{\vspace*{-1pt}% 
\noindent{\color{cream}\rule[-\figrulesep]{\textwidth}{1.5pt}} }

\makeatother
%%%END OF FIGURE SETUP%%%

%%%TITLE, AUTHORS AND ABSTRACT%%%
\twocolumn[
  \begin{@twocolumnfalse}
\vspace{3cm}
\sffamily
\begin{tabular}{m{4.5cm} p{13.5cm} }

\includegraphics{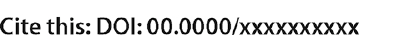} & \noindent\LARGE{In-situ investigation of temperature induced agglomeration in non-polar magnetic nanoparticle dispersions by small angle X-ray scattering} \\%Article title goes here instead of the text "This is the title"
\vspace{0.3cm} & \vspace{0.3cm} \\

 & \noindent\large{Christian Appel,$^{\ast}$\textit{$^{a,b}$} Bj\"orn Kuttich,\textit{$^{a,c}$} Tobias Kraus\textit{$^{c,d}$} and Bernd St\"uhn\textit{$^{a}$}} \\%Author names go here instead of "Full name", etc.

\includegraphics{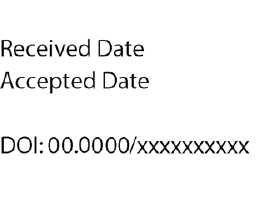} & \noindent\normalsize{Non-polar magnetic nanoparticles agglomerate upon cooling. This process is followed by in-situ small angle X-ray scattering to assess structural properties of the emerging agglomerates. On the length scale of a few particle diameters, no differences are found between the agglomerates of small ($d=\SI{12}{\nano\metre}$) and large ($d=\SI{22}{\nano\metre}$) nanoparticles. Hard-sphere like random packing with a local packing fraction of $\eta=0.4$ is seen. On larger length scales, small particle form compact superstructures, while large particles arrange into agglomerates that resemble chain-like structure in SAXS. This can be explained by directed magnetic dipole interactions that dominate larger particles, while isotropic van der Waals interaction governs the agglomeration of smaller particles.} \\%The abstrast goes here instead of the text "The abstract should be..."

\end{tabular}

 \end{@twocolumnfalse} \vspace{0.6cm}

  ]
%%%END OF TITLE, AUTHORS AND ABSTRACT%%%

%%%FONT SETUP - please do not change any commands within this section
\renewcommand*\rmdefault{bch}\normalfont\upshape
\rmfamily
\section*{}
\vspace{-1cm}

%%%FOOTNOTES%%%

\footnotetext{\textit{$^{a}$~Experimental Condensed Matter Physics, TU Darmstadt, Germany; $^{b}$~current affiliation: Paul Scherrer Institut, 5232 Villigen PSI, Switzerland.  E-mail: christian.appel@psi.ch}; \textit{$^{c}$~INM - Leibniz Institute for New Materials, Campus D2 2, 66123 Saarbr\"{u}cken, Germany}; \textit{$^{d}$~Colloid and Interface Chemistry, Saarland University, Campus D2 2, 66123 Saarbr\"{u}cken, Germany}}
%\footnotetext{\textit{$^{c}$~INM - Leibniz Institute for New Materials, Campus D2 2, 66123 Saarbr\"{u}cken, Germany },}

%Please use \dag to cite the ESI in the main text of the article.
%If you article does not have ESI please remove the the \dag symbol from the title and the footnotetext below.
%\footnotetext{\dag~Electronic Supplementary Information (ESI) available: [details of any supplementary information available should be included here]. See DOI: 00.0000/00000000.}
%additional addresses can be cited as above using the lower-case letters, c, d, e... If all authors are from the same address, no letter is required

%\footnotetext{\ddag~Additional footnotes to the title and authors can be included \textit{e.g.}\ `Present address:' or `These authors contributed equally to this work' as above using the symbols: \ddag, \textsection, and \P. Please place the appropriate symbol next to the author's name and include a \texttt{\textbackslash footnotetext} entry in the the correct place in the list.}

%%%END OF FOOTNOTES%%%

%%%MAIN TEXT%%%%
\section*{Introduction}

Metal nanoparticles stabilised by non-polar ligands have many interesting properties due to their unique combination of strongly attracting cores and complexly interacting shells.\citep{Yang2016,Kister2018,Doblas2019,Balan2019,Heuer2019,Monego2020} Their colloids are used as conductive inks, inkjet-printed to form sensors \citep{Singh2010,Jensen2011} or functional components in sensors\citep{Shipway2000,Saha2012} and photovoltaics.\citep{Talapin2005} Non-polar nanoparticles are fundamentally interesting because the simple and elegant DLVO theory is unable to explain their colloidal stability due to the missing charge stabilisation. Stability depends on  ligand/ligand and ligand/solvent interactions, which have to counteract the strong attractive forces between metal nanoparticles.\cite{Abecassis2008,Khan2012,Kister2018} Recent studies focused on non-polar noble metal (Ag and Au) or semiconductor (CdSe and CdS) particles, for which attraction is mainly based on the isotropic van der Waals force.\citep{Yang2016,Kister2018,Monego2018,Doblas2019,Balan2019,Heuer2019,Monego2020} Very little is known on the colloidal stability of magnetic nanoparticles in the absence of an external magnetic field. Depending on particle size, precise composition and crystal structure, either magnetic dipole or van der Waals interactions can induce their agglomeration.

The temperature-dependant stability of non-polar nanoparticles is a sensitive and accessible probe for their interactions. Agglomeration upon cooling below well-defined temperatures was found for particles with cores of Au and CdSe. The precise agglomeration temperature depended on particle size, ligand length and structure as well as the solvent.\citep{Kister2018,Monego2018,Monego2020} Agglomeration was dominated by interactions between shells or cores depending on particle and ligand sizes. Shell-induced agglomeration was dominated by a disorder-order phase transition of the ligand shell, while core-dominated agglomeration was due to van der Waals interactions of the metal cores that overcame steric repulsion of the ligands by more than the thermal energy k$_BT$.\citep{Roke2006,Abecassis2008,Khan2012,Kister2018,Monego2018,Monego2020}

This contribution analyses the temperature-dependent agglomeration of magnetic iron oxide nanoparticles (FeNP). These particles interact not only via isotropic van der Waals attraction but also via highly directed magnetic dipole interactions. In order to focus on this specific property of the nanoparticle core, we investigated relatively large particles ($d\geq\SI{10}{\nano\metre}$) stabilised by oleic acid, an unsaturated acid for which no disorder-order transition of the ligand shell is expected. We expect agglomeration to be fully core-dominated. We first studied the dispersions by small angle X-ray scattering (SAXS) at room temperature where no agglomeration is seen. We then followed the agglomeration process upon cooling in-situ with SAXS revealing distinct differences in the agglomerate structure for different particles sizes. These findings were finally correlated with the magnetic properties of the single particles leading to a conclusive explanation based on competing van der Waals and magnetic interactions. 

\begin{table*}[h]
\small
  \caption{\ Structural and magnetic properties of the investigated iron oxide nanoparticles}
  \label{tab:properties}
  \begin{tabular*}{\textwidth}{@{\extracolsep{\fill}}lllllll}
    \hline
    sample ID & $d_{\text{core}}$ / \SI{}{\nano\meter} & PDI$_{\text{core}}$ & $m_{\text{S}}$ per particle / \SI{}{\milli\ampere\nano\meter\squared}  & $m_{\text{R}}$ / $m_{\text{S}}$  \\
    \hline 
    FeNP$_{10}$ & 11.8 & 0.11& 134& 0.05 \\
    FeNP$_{20}$ & 21.5 & 0.11& 625& 0.09 \\
    \hline
  \end{tabular*}
\end{table*}

\begin{figure*}[t]
 	\includegraphics[width=18.2cm]{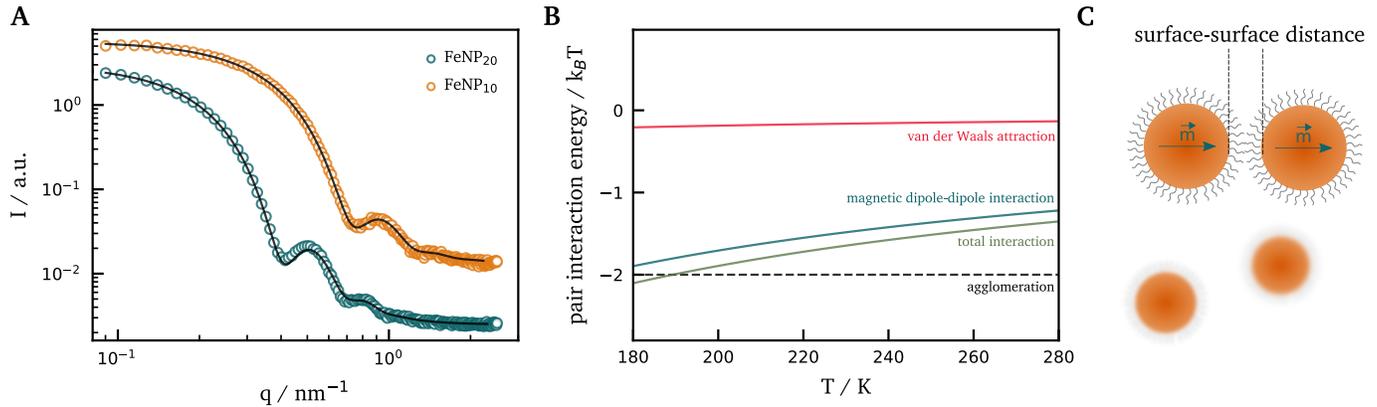}  
 	\caption{\ \textbf{A} Small-angle scattering curves of FeNP$_{10}$ and FeNP$_{20}$ at \SI{293}{\kelvin}. Solid lines are fits of a polydisperse spherical form factor model. \textbf{B} Estimated pair particle interaction energies for FeNP$_{20}$ calculated for the particle diameter and magnetic moment given in table \ref{tab:properties}. Assuming a surface-surface distance of \SI{4}{\nano\metre} and a head to tail configuration of the magnetic moments as indicated in \textbf{C}.} 
 	\label{fig:properties}
\end{figure*}

\section*{Experimental section}

Spherical iron oxide nanoparticles were obtained from \textit{Ocean Nanotech} (sample ID: FeNP$_{10}$ \& FeNP$_{20}$, $d \approx \SI{10}{}$ and $\SI{20}{\nano\meter})$.  The dispersion in chloroform was stabilised by an oleic acid shell of roughly \SI{2}{\nano\meter} thickness. According to the manufacturer, the nanoparticles were composed of \ce{Fe3O4} nanocrystals. The magnetic moments were measured by a vibrating sample magnetometer using magnetic hysteresis loops at room temperature. The resulting saturation magnetisation $m_{\text{S}}$ and remanent magnetisation $m_{\text{R}}$ per nanoparticle are given in table \ref{tab:properties} for FeNP$_{10}$ and FeNP$_{20}$. Details on the investigation can be found in a previous study.\cite{Appel2019}

Small Angle X-ray Scattering (SAXS) experiments were carried out using the K$_\alpha$-line ($\lambda = \SI{0.154}{\nano\metre}$) of a conventional Copper anode with a laboratory X-ray set-up. 
The scattered intensity of the point focused beam, collimated by 3 pinholes, was measured at a sample-detector distance of \SI{1.5}{\meter} with a two-dimensional multiwire gas detector (Molecular Metrology, 1024 x 1024 pixels).
The scattering vector $\vec{q}$, $q=\left|\vec{q}\right| = 4\pi/\lambda\sin(\theta)$ with the scattering angle 2$\theta$ was calibrated by measuring silver behenate.\cite{Huang1993} An overall instrumental resolution of $\Delta q = \SI{0.1}{\nano\metre}^{-1}$ was estimated from the width of the first diffraction peak.
Temperature dependent measurements ranging between \SI{153}{\kelvin} and \SI{293}{\kelvin} were performed on samples in borosilicate capillaries, \SI{1.5}{\milli\meter} diameter, inserted into a Linkam stage in the evacuated sample chamber.
Each temperature was approached with a cooling rate of \SI{20}{\kelvin\per\minute} and the sample was given an additional \SI{15}{\minute} for equilibration.
Scattering curves were recorded every \SI{10}{\kelvin} for decreasing temperature with an exposure time of \SI{30}{\minute}. All scattering was isotropic, data was radially averaged and the accessible $q$-range is  $\SI{0.07}{\nano\metre}^{-1} \leq q \leq \SI{2.5}{\nano\metre}^{-1}$.

\section*{Results and discussion}

\subsection*{Dispersion stability at room temperature}

As-received nanoparticles are dispersed in chloroform which is not suited for X-ray scattering because of the strong X-ray absorption of chlorine. Solvent exchange was therefore performed by complete evaporation of chloroform and re-dispersing the particles in toluene. The concentration was set to \SI{10}{\milli\gram\per\milli\liter}. In this first section we investigate structural properties of both systems at room temperature by SAXS to determine precise size distributions and ensure the absence of any agglomerates.

X-ray scattering comprises information on the variation of electron density within the samples and the intensity measured by the detector is 
\begin{equation}
	\text{I}(q) \sim \langle\text{F}(q)\rangle^2\cdot\text{S}(q)
	\label{eq:SAXS_int}
\end{equation}
with the form factor $\text{F}(q)$ and the structure factor $\text{S}(q)$.
For a dilute system of non interacting particles, the structure factor $\text{S}(q)$ becomes 1 and size as well as shape of the particles can be deduced from the form factor.\cite{Guinier1963}

Figure \ref{fig:properties} \textbf{A} shows the scattering profiles of FeNP$_{10}$ and FeNP$_{20}$ with a constant background contribution from toluene.
The low contrast difference between toluene and oleic acid in X-rays makes it impossible to differentiate shell and solvent, thus only the core of the particles is considered.
At small $q$, the scattering curves exhibit a plateau before intensity decays. 
The characteristic oscillations combined with an intensity decay of $\text{I} \sim q^{-4}$ are typical for a form factor of spherical particles, which is able to nicely describe scattering from both samples (black lines). The absence of any structure factor contribution to the scattering signal proves that all particles are well dispersed, no agglomerates can be observed. Core size and polydispersity (PDI), i.e. relative width of the size distribution, can be deduced for each sample and the obtained values are summarised in table \ref{tab:properties}.

The pair interaction energies between two nanoparticles can be estimated based on the particle sizes and magnetic moments from above (fig.\ref{fig:properties} \textbf{B}). Thickness of the oleic acid shell is estimated based on the molecule's structure to be \SI{2}{\nano\metre},\cite{Borges2011} we therefore calculated interaction energies for a particle surface-surface separation of $s=\SI{4}{\nano\metre}$. The ligand molecules do not interdigitate at this spacing and thus do not contribute significantly to the nanoparticle interaction. A Hamaker constant of $A_{131}=\SI{9e-21}{\joule}$ was used for \ce{Fe3O4} interacting across toluene.\cite{Faure2011} The maximal magnetic dipole-dipole interaction was estimated by assuming a head to tail configuration of the magnetic moments as in figure \ref{fig:properties} \textbf{C}. The resulting magnetic interaction at room temperature was $-1.1\ \text{k}_BT$, which exceeds the van der Waals attraction of $-0.3\ \text{k}_BT$. Assuming that agglomeration occurs if the pair particle interaction energy exceeds -2 k$_BT$, an agglomeration temperature of \SI{190}{\kelvin} can be estimated. This does not consider ligand/solvent and ligand/ligand interactions that affect nanoparticle agglomeration.\cite{Kister2018}

\subsection*{Temperature induced agglomeration of FeNPs}

\begin{figure*}
	\centering
 	\includegraphics[width=18.2cm]{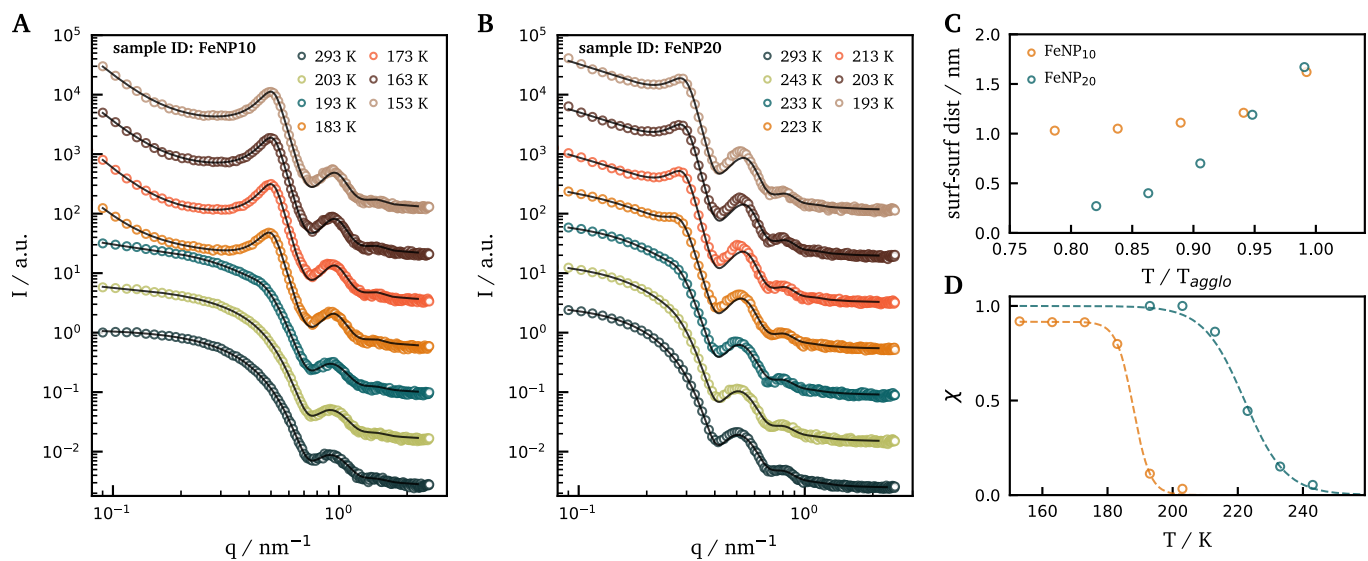} 
 	\caption{\ \textbf{A} Small-angle scattering curves of FeNP$_{10}$ measured in a temperature range of \SI{153}{\kelvin} to \SI{293}{\kelvin} (cooling protocol in main text). Solid lines represent fits using equation \ref{eq:SAXS_int} and \ref{eq:SAXS_HS}. \textbf{B} Scattering data measured for FeNP$_{20}$ in a temperature range of \SI{193}{\kelvin} to \SI{293}{\kelvin}. \textbf{C} Structural properties obtained from model fits, such as average particle surface-surface distance and agglomeration fraction $\chi$ (\textbf{D}) as obtained for both particle sizes and  all temperatures.}
 	\label{fig:agglomeration}
\end{figure*}

The onset of agglomeration upon decreasing temperature was studied for both systems. We investigated the small angle scattering from FeNP$_{10}$ and FeNP$_{20}$ dispersed in toluene from \SI{293}{\kelvin} to \SI{153}{\kelvin}.
Well-dispersed FeNP were characterized with SAXS at \SI{293}{\kelvin} as discussed in the previous section. We then reduced the sample temperature in steps of \SI{10}{\kelvin} at a cooling rate of \SI{20}{\kelvin\per\minute} and recorded scattering at each step after an equilibration time of \SI{15}{\minute} during an exposure time of \SI{30}{\minute}.
At the end of the cycle, samples were heated to \SI{293}{\kelvin} and remeasured after \SI{1}{\hour}.
This temperature cycle was repeated for at least three different samples and yielded almost identical scattering patterns at all temperatures.
Selected SAXS curves of FeNP$_{10}$ and FeNP$_{20}$ are presented in Figure \ref{fig:agglomeration} \textbf{A} and \ref{fig:agglomeration} \textbf{B}. Along with the \SI{298}{\kelvin} measurement only data which shows a significant structure factor contribution is shown.
A distinct maximum appeared at temperatures below \SI{193}{\kelvin} and \SI{233}{\kelvin} for small and large particles respectively ($q \approx \SI{0.5}{\nano\metre}^{-1}$ for FeNP$_{10}$ and $q \approx \SI{0.3}{\nano\metre}^{-1}$ for FeNP$_{20}$) accompanied by an increase in intensity for small $q$.
This emerging feature is related to the structure factor and typically observed for agglomerating nanoparticles.\cite{Kister2018}
Both features vanished once the sample was reheated to \SI{293}{\kelvin}, indicating that the agglomeration is reversible.

Agglomeration did not occur at a sharply defined temperature, as is particularly obvious for FeNP$_{20}$. The temperature range of the transition is presumably caused by a distribution of particle sizes and ligand densities, leading to slightly different agglomeration temperatures for individual nanoparticles. We modelled the agglomeration process using a modified version of equation \ref{eq:SAXS_int}. At any given temperature a certain fraction of particles $\chi_\text{agglo}$ is part of an agglomerate, while $1-\chi_\text{agglo}$ is the fraction of freely dispersed particles. Scattering from both fractions adds up incoherently, leading to the following ansatz for the total scattered intensity:
\begin{equation}
	\text{I}(q) \sim \langle\text{F}(q)\rangle^2\cdot\left[ (1-\chi_\text{agglo}) + \chi_\text{agglo} \cdot S(q)\right]
	\label{eq:SAXS_HS}
\end{equation}
The first part of the sum describes pure form factor scattering from dispersed particles, while the second part stems from the combined form and structure factor scattering of agglomerates. The agglomeration fraction $\chi_\text{agglo}$ weighs the contributions. Fitting requires an appropriate structure factor model. The dominant features in the experimentally observed scattering patterns that can be attributed to the structure factor are a single pronounced peak at low $q$-values and a power-law like scattering at even lower $q$. Higher order structure peaks are absent, which indicates randomly packed particles in the agglomerates. We describe them using the Hard-Sphere structure factor $S_\text{HS}(q)$ \citep{Percus1958,Thiele1963} combined with a phenomenological power-law relation to describe scattering at low $q$-values:
\begin{equation}
	S(q) = S_\text{HS}(q)+c\cdot q^{-\nu}
\label{eq:SAXS_sq}
\end{equation}

We first fitted the dispersed samples at \SI{293}{\kelvin} to determine the two main form factor parameters, particle diameter $d$ and its relative polydispersity PDI (table \ref{tab:properties}). Both are kept constant in all fits for lower temperatures, while the structure factor parameters, hard-sphere volume fraction $\eta$, hard-sphere diameter $d_\text{HS}$, power-law pre-factor $c$ and power-law exponent $\nu$ are fitted together with the agglomeration fraction $\chi_\text{agglo}$. The two structure factor contributions are well separated in $q$-space, thus the different fitting parameters do not correlate and can be precisely determined for each data-set. Note that the hard-sphere volume fraction is not the overall particle volume fraction but indicates particle density within the agglomerates.

Figures \ref{fig:agglomeration} \textbf{A} and \textbf{B} indicates that the model fits the data well. Deviations at the lowest temperatures around the form factor maximum are likely due to the monodisperse structure factor approximation of equation \ref{eq:SAXS_int}. Figure \ref{fig:agglomeration} \textbf{C} indicates the deduced agglomeration fractions $\chi_\text{agglo}$ as a function of temperature that follow a sigmoidal curve for both particle sizes upon cooling. We define the temperature at which $\chi_\text{agglo}=0.2$ as "agglomeration temperature"\cite{Kister2018} and find $T_\text{agglo}=\SI{192}{\kelvin}$ and $T_\text{agglo}=\SI{231}{\kelvin}$ for FeNP$_{10}$ and FeNP$_{20}$ respectively. The temperature range of the agglomeration transition varied between  \SI{8}{\kelvin} ($T(\chi_\text{agglo}=0.2)-T(\chi_\text{agglo}=0.8)$) for the smaller particles and \SI{25}{\kelvin} for the larger particles. We attribute this to the higher absolute polydispersity of the FeNP$_{20}$ system.

The model lets us explore differences in the structure factor parameters for the two particle sizes. Agglomerates of both particle systems had a hard-sphere volume fraction of $\eta\approx0.4$, which is considerably lower than the theoretical maximum value for monodisperse hard-spheres of about 0.64.\cite{Torquato2000} We used it to calculate the average particle surface-surface distance in \ref{fig:agglomeration} \textbf{D} as a function of temperature relative to agglomeration temperature. The surface-surface distance of both particle types just below agglomeration temperature was \SI{1.6}{\nano\metre}, indicating strongly interdigitated ligand shells. Particle spacing dropped upon lowering temperature, presumably because the steric repulsion of the ligand molecules scales with k$_BT$ while magnetic and van der Waals attraction hardly depend on temperature.\cite{Kister2018} Since the attractive interactions are significantly stronger for the larger particles, a much smaller spacing is found for the FeNP$_{20}$ system for similar under-cooling below $T_\text{agglo}$.

Considering the power-law feature at the lowest measured q-values that is visible in figure \ref{fig:agglomeration} \textbf{A} and \textbf{B}, its intensity increases with decreasing temperature, and the increase is more pronounced for the smaller particles. This change may be connected to a more compact structure of the largest agglomerates. Power-law exponents in the range of 3 for FeNP$_{10}$ and around unity for FeNP$_{20}$ suggest that small particles form compact agglomerates, while agglomerates of large particles are more open, even linear. Isotropic van der Waals attractions (that scale with the square of particle distance) will favour compact agglomerates, while the highly directional magnetic dipole-dipole interactions (that scale with the distance cubed) should lead to a chain-like conformation of the particles. A similar trend was found for smaller particles in a 2D arrangement at the air/water interface,\cite{Appel2019} while chain like conformations of \ce{Fe2O3} nanocrystalls were claimed to exist already for \SI{7}{\nano\metre} sized particles.\cite{Bonini2007}

\section*{Conclusions}

The agglomeration of non-polar iron oxide nanoparticles upon temperature decrease was investigated. Particles were covered by a shell of unsaturated oleic acid which does not exhibit a disorder-order phase transition and thus stabilises nanoparticles efficiently in a large temperature range. Low agglomeration temperatures of $T_\text{agglo}=\SI{192}{\kelvin}$ and $T_\text{agglo}=\SI{231}{\kelvin}$ were found for particles with a diameter of \SI{11.8}{\nano\meter} and \SI{21.5}{\nano\meter}. The experimentally determined magnetic moments per particle were $m_{\text{S,FeNP}_{10}} =\SI{134}{\milli\ampere\square\nano\meter}$ and $m_{\text{S,FeNP}_{20}}=\SI{625}{\milli\ampere\square\nano\meter}$; they decrease the dispersion stability by adding an additional attractive interaction on the order of $1.5\ \text{k}_BT$ per particle at $T_\text{agglo}$.

The interplay of two different attractive core-core interactions affected the agglomerate structures as a function of core diameter. Local particle packing in the agglomerates was hard-sphere like and random at a packing fraction around $\eta=0.4$ for both cases, but distinct differences emerged at larger length scales. Agglomerates of small particles were compact, while large particles agglomerated into more open, linear structures. Both findings differ significantly from agglomerate structures found for field-induced agglomeration of similar particles, where formation of a FCC supercrystal was seen.\cite{Fu2016} Fu et al.\ proposed a two-step formation mechanism where dispersed particles first form 1D chains aligned parallel to the external field. The chains then arrange into a 3D FCC supercrystal.\cite{Fu2016}

We estimated the ratio between maximal magnetic and van der Waals interactions in our system as 1.6 for the small and 2.7 for the large particles, indicating that oriented magnetic dipole-dipole interactions exceeded isotropic van der Waals attraction for both cases. However, only for the large particles magnetic interaction can dominate the agglomeration leading to the formation of particle chains. In the absence of an external magnetic field these chains have no preferred orientation, which strongly hinders the formation of a 3D supercrystal as proposed by Fu et al.\cite{Fu2016} Particle chains cannot align parallel efficiently because of their presumably slow rotational diffusion. A lose network of unoriented chains may be formed resembling a randomly packed structure in small angle X-ray scattering.

\section*{Conflicts of interest}
There are no conflicts to declare.

%\section*{Acknowledgements}
%The Acknowledgements come at the end of an article after Conflicts of interest and before the Notes and references.

%%%END OF MAIN TEXT%%%


\balance


%%%REFERENCES%%%
\bibliography{Literature} %You need to replace "rsc" on this line with the name of your .bib file
\bibliographystyle{rsc} %the RSC's .bst file

\end{document}